# A Unified Linear Precoding Design for Multi-user MIMO Systems

Md. Abdul Latif Sarker

*Abstract*—We address the problem of the bit-error-rate (BER) performance gap between the sub-optimal and optimal linear precoder (LP) for a multi-user (MU) multiple-input and multiple-output (MIMO) broadcast systems in this paper. Particularly, mobile users suffer noise enhancement effect due to a sub-optimal LP that can be suppressed by an optimal LP matrix. A sub-optimal LP matrix such as a linear *zero-forcing* (LZF) precoder performs in high signal-to –noise-ratio (SNR) regime only, in contrast, an optimal precoder for instance a linear *minimum mean-square-error* (LMMSE) precoder outperforms in both low and high SNR scenarios. These kinds of precoder illustrates the BER gap distance at least 0.1 when it is used in itself in a MU-MIMO systems. Thus, we propose and design a unified linear precoding (ULP) matrix using a precoding selection technique that combines the sub-optimal and optimal LP matrix for a multi-user MIMO systems to ensure zero BER performance gap in this paper. The numerical results show that our proposed ULP technique offers significant performance in both low and high SNR scenarios.

## I. INTRODUCTION

PRECODING is an important technique to explore the significant performance in terms of bit-error rate (BER) as well as the achievable sum-rates for MU-MIMO downlink transmission [1-2]. The most common linear precoding scheme such as a LZF precoding in [3-4], a LMMSE precoding in [1, 4] and the nonlinear precoding like dirty-paper coding (DPC) based precoding in [4-5].

Recently, a hybrid precoding scheme has proposed in [1-2, 6-8]. In [1], authors shown a multi-stage robust hybrid linear precoding in a multi-user MIMO systems and proposed the several kind of MMSE precoding based two efficient iterative algorithms. In [2], Authors considered low-complexity hybrid precoding in the massive multi-user MIMO systems and proposed the full-complexity ZF linear precoding to enhance the spectral efficiency of the systems. In [6-7], authors shown a two-tier hybrid precoder scheme to approach the performance of the traditional LZF precoder in a multi-user massive MIMO system. There are many papers on LZF and LMMSE precoding focusing on different design criteria in [9-12]. Authors in [8], also shown a two-tier precoder for block diagonalization of a multiuser MIMO channel including other-cell interference.

All of the above related works have considered the traditional LZF and LMMSE based precoding. Usually, the conventional LZF precoder achieves the performance close to the sum-capacity when the system interference is limited or the number of users become large, otherwise, it requires significant feedback overhead with respect to SNR while an imperfect CSI at the transmitter that provides the significant throughput loss due to residual multiuser interference. In addition, the conventional LMMSE precoder does not work properly when the user is qualify large in order to a multi-user interference environment. However, we thus far notice that the BER performance gap between the traditional LZF and LMMSE precoder is still high as in [1-4, 6-13].Thus, we propose and design a unified linear precoding (ULP) scheme that overcome the BER performance gap completely in this paper.

This paper is organized as follows:
First, we investigate the system model and problem formulation in Section II. Then we design a ULP in Section III. Finally, numerical results and conclusions are presented in Section IV and Section V.

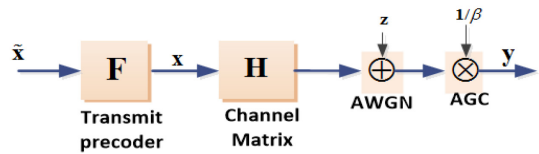

Fig. 1. Linear Precoding schemes in [4].

## II. SYSTEM MODEL AND PROBLEM FORMULATION

### A. System Model

We consider a multiuser MIMO system with downlink channel $\mathbf{H} \in \mathbb{C}^{K_t M_{R,k} \times M_T}$ whereas the total $K_t$ users each employing $M_{R,k}$ receive antennas for $k-th$ $(k=1,2,...,K)$ receivers and receiving their own data streams are precoded transmit symbol vector $\mathbf{x} \in \mathbb{C}^{M_T \times 1}$ for $K_t$ users that can be

Md. Abdul Latif Sarker is with the Chonbuk National University, Jeonju, 54896 Korea (e-mail: latifsarker@jbnu.ac.kr).

expressed as $\mathbf{x} = \mathbf{F}\tilde{\mathbf{x}}$ at the BS with $M_T$ transmitting antennas where $\mathbf{F}$ is the $M_T \times M_T$ LP matrix and $\tilde{\mathbf{x}}$ is the original symbol vector for transmission, respectively in Fig.1. Then, the received signal vector $\mathbf{y}$ for all user is given by

$$\mathbf{y} = \mathbf{HF}\tilde{\mathbf{x}} + \mathbf{z} \tag{1}$$

where $\mathbf{z} \in \mathbb{C}^{K_t M_{R,k} \times 1}$ is an additive white Gaussian noise (AWGN) vector with noise covariance matrix $\mathbb{E}\{\mathbf{zz}^H\} = N_0 \mathbf{I}_{K_t M_{R,k}}$, the operator $(\cdot)^H$ is called Hermitian, $N_0$ is the noise variance, and $\mathbf{I}_{K_t M_{R,k}}$ is the $K_t M_{R,k} \times K_t M_{R,k}$ identity matrix, respectively.

B. *Problem Formulation*

Let, the traditional LP matrix, $\mathbf{F}$ is an optimal LP matrix, that is given by [4, 1-2, 13-14]

$$\mathbf{F} = \beta \mathbf{H}^H \left(\mathbf{HH}^H + m\sigma^2 \mathbf{I}\right)^{-1}$$
$$= \begin{cases} \mathbf{F}_{LZFP} = \beta_{LZFP} \mathbf{H}^H \left(\mathbf{HH}^H\right)^{-1}, & if\ m=0; \\ \mathbf{F}_{LMMSEP} = \beta_{LMMSEP} \mathbf{H}^H \left(\mathbf{HH}^H + m\sigma^2 \mathbf{I}\right)^{-1}, & if\ m>0. \end{cases} \tag{2}$$

where $m \geq 0$ is a constant that indicates the channel inversion or the regularized channel inversion, and $\beta$ is a constant to meet that the total transmitted power constraint after precoding and it is given as

$$\beta = \sqrt{\frac{M_T}{Tr(\mathbf{FF}^H)}}. \tag{3}$$

Thus, the estimated signal $\hat{\tilde{\mathbf{x}}}$ after a conventional LP for all user is given by

$$\hat{\tilde{\mathbf{x}}} = \beta^{-1}\mathbf{y} = \beta^{-1}\left(\mathbf{HF}\tilde{\mathbf{x}} + \mathbf{z}\right) \tag{4}$$

In reality, we observe that in (2), if m=0 or $m>0$, the LP matrix $\mathbf{F}$ indicates the characteristic of the LZF precoding or the LMMSE precoding and the BER performance demonstrates a very high gap between LZF and LMMSE precoding schemes. Thus, we design a unified linear precoding (ULP) to mitigate this problem in next Section III.

### III. PROPOSED ULP MATRIX DESIGN

We consider a unified channel matrix $\mathbf{H}_u$ is

$$\mathbf{H}_u = \begin{bmatrix} \mathbf{H} & u\mathbf{I} \end{bmatrix}^T \tag{5}$$

where $u \geq 0$ is a constant that indicates individual or unified precoding, and the *pseudo-inverse* of a unified channel matrix $\mathbf{H}_u$ is given by [14]

$$\mathbf{H}_u^\dagger = \mathbf{H}_u^H \left(\mathbf{H}_u \mathbf{H}_u^H\right)^{-1}. \tag{6}$$

By setting (5) in (2) and design a ULP to combine the LZF and LMMSE precoding is as follows:

$$\mathbf{F}_u = \beta_u \mathbf{H}_u^H \left(\mathbf{H}_u \mathbf{H}_u^H + m\sigma^2 \mathbf{I}\right)^{-1}$$
$$= \begin{cases} \mathbf{F}_{LZFP} = \beta_{LZFP} \mathbf{H}^H \left(\mathbf{HH}^H\right)^{-1}, & if\ u=0; m=0; \\ \mathbf{F}_{LMMSEP} = \beta_{LMMSEP} \mathbf{H}^H \left(\mathbf{HH}^H + m\sigma^2 \mathbf{I}\right)^{-1}, & if\ u=0; m>0; \\ \mathbf{F}_{ULZFP} = \beta_{UZFP} \mathbf{H}_u^H \left(\mathbf{H}_u \mathbf{H}_u^H\right)^{-1}, & if\ u>1; m=0 \\ \mathbf{F}_{ULMMSEP} = \beta_{UMMSEP} \mathbf{H}_u^H \left(\mathbf{H}_u \mathbf{H}_u^H + m\sigma^2 \mathbf{I}\right)^{-1}, & if\ u>0; m>0; \end{cases} \tag{7}$$

where $\beta_u$ is an estimated unified constant as in (3), that is

$$\beta_u = \sqrt{\frac{M_T}{Tr(\mathbf{F}_u \mathbf{F}_u^H)}}. \tag{8}$$

However, to compensate for the effect of amplification by a factor of $\beta_u$ at the transmitter, the received signal must be divided by $\beta_u$ via automatic gain control (AGC) at the receiver as depicted in Fig.1. Thus, the estimated signal $\hat{\tilde{\mathbf{x}}}_u$ after a unified precoding for all users is given by

$$\hat{\tilde{\mathbf{x}}}_u = \beta_u^{-1} \mathbf{y}_u = \beta_u^{-1}\left(\mathbf{H}_u \mathbf{F}_u \tilde{\mathbf{x}}_u + \mathbf{z}_u\right)$$
$$= \beta_u^{-1} \mathbf{H}_u \beta_u \mathbf{H}_u^H \left(\mathbf{H}_u \mathbf{H}_u^H + m\sigma^2 \mathbf{I}\right)^{-1} \tilde{\mathbf{x}}_u + \tilde{\mathbf{z}}_u, \tag{9}$$

where $\mathbf{y}_u = \begin{bmatrix} \mathbf{y} & 0 \end{bmatrix}^T$, $\tilde{\mathbf{x}}_u = \begin{bmatrix} \tilde{\mathbf{x}} & 0 \end{bmatrix}^T$, $\tilde{\mathbf{z}}_u = \beta_u^{-1} \mathbf{z}_u$ $\mathbf{z}_u = \begin{bmatrix} \mathbf{z} & -c\tilde{\mathbf{x}} \end{bmatrix}^T$, respectively.

### IV. NUMERICAL RESULTS

We numerically compare our designed ULP scheme in Fig.2 to Fig.4 against the conventional LP scheme for a multiuser MIMO broadcast systems. In computer simulations, we consider $M_T = 8$, $M_{R,k} = 1$, $K_t = 20$ and assume the total active users, $K_{at} = M_T = 8$ in which $K_{at} = 8$ users with the highest norm values are selected out of $K_t = 20$. The quadrature phase shift keying (QPSK) modulation scheme is used for the symbol normalizing at 8 transmit antennas with 10 frames. In the independent and identically distributed (i.i.d) Rayleigh fading channel environment, we treat 1,000 times of Monte Carlo channel realizations in the computer simulations.

In this paper, Fig.2 to Fig.4 shows the BER performance gap between the conventional LP and proposed ULP schemes in at several SNR values such as 14, 20 and 30 [dBs] SNR values. The measured performance BER gap is $1.1 \times 10^{-1}$ in Table I between the LZF and LMMSE precoding at 14 [dBs] SNR values when m=0 and m=1 is applied in (2) which has been shown in Fig.2. In contrast, we observe that, the zero BER gap between the LZF and LMMSE precoding based on the proposed ULP scheme when u=0, m=0 and u=0, m>0 is applied in (7). The proposed ULP schemes outperform the slightly BER gap between ULZF and ULMMSE precoding at 14 [dBs] SNR cases only if u=1, m=0 and u=1, m=1 is applied in (7) and Table I.

Similarly, in Fig.3 to Fig.4 illustrates $4.7 \times 10^{-2}$ and

$5.3\times10^{-3}$ BER performance gap between the traditional LZF and LMMSE precoding at 20 [dBs] and 30 [dBs] SNR values respectively. Looking in Fig.2 to Fig.4, we note that the channel gains of almost 1.99, 3.24 and 3.63[dBs] are lost using the traditional LP schemes at 14, 20 and 30 [dBs] SNR scenarios, respectively.

We also observe in Table I, there is no BER performance gap of LZF, LMMSE, ULZF and ULMMSE while a unified precoder select u=0, m=0 and u=0, m=1 at high SNR scenarios but the low SNR scenario illustrates a negligible BER gap such as $1.5\times10^{-3}$ BER gap (in Table I) between the designed ULZF and ULMMSE precoding when a ULP is selected u=1, m=0 and u=1, m=1, respectively. However, the designed ULP schemes ensure that a zero BER performance gap between the sub-optimal and optimal precoding that also confirm high channel gains regarding the numerical and simulation results in this paper.

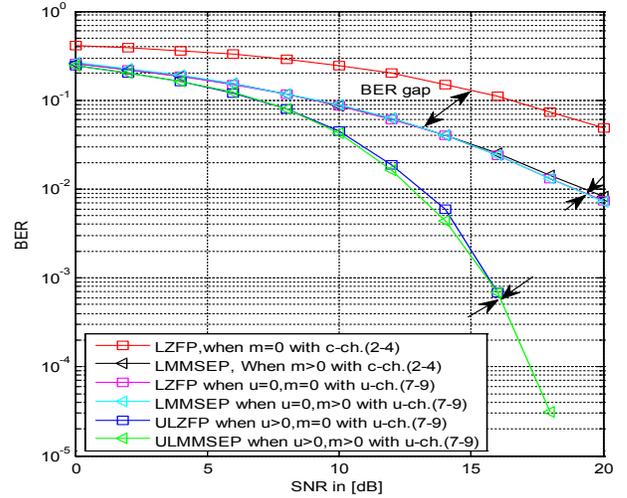

Fig.3. BER performance gap of LP and ULP schemes at 20 [dBs] SNR.

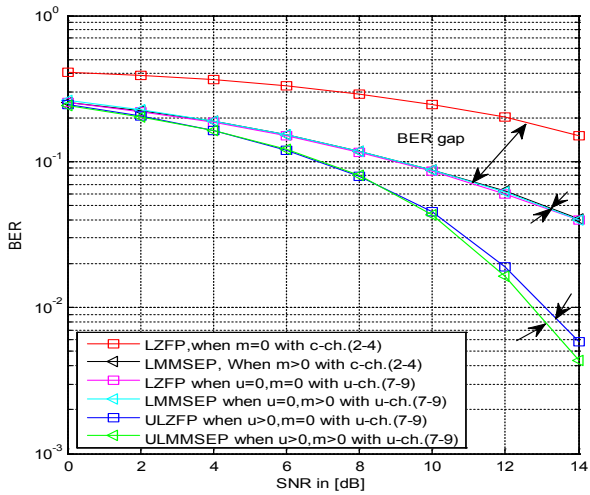

Fig.2. BER performance gap of LP and ULP schemes at 14 [dBs] SNR.

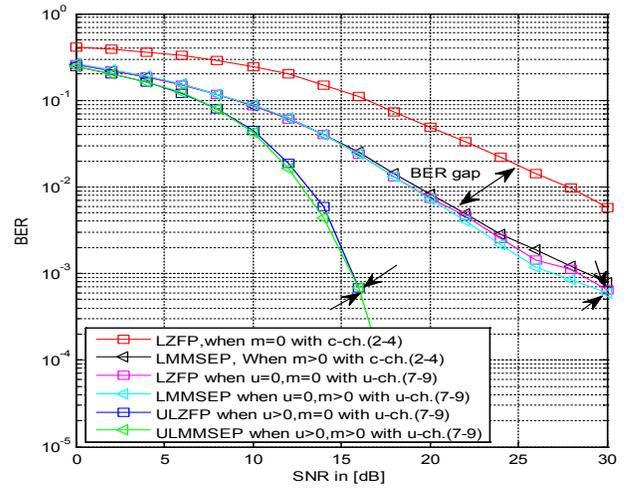

Fig.4. BER performance gap of LP and ULP schemes at 30 [dBs] SNR.

TABLE I
THE PERFORMANCE GAP OF SEVERAL SNR CASES BETWEEN THE LP AND ULP SCHEMES.

| SNR in [dB] | Conventional channel (c-ch.) based LP scheme in (2-4) | | | Proposed unified channel (u-ch.) based ULP scheme in (7-9) | | | | | |
|---|---|---|---|---|---|---|---|---|---|
| | LZFP (m=0) | LMMSEP (m>0) | BER Gap | LZFP (u=0,m=0) | LMMSEP (u=0, m>1) | BER Gap | ULZFP (u>0,m=0) | ULMMSEP (u>0, m>0) | BER Gap |
| 14 | $1.5\times10^{-1}$ | $4.0\times10^{-2}$ | $1.1\times10^{-1}$ | $4.0\times10^{-2}$ | $4.0\times10^{-2}$ | 0 | $6.0\times10^{-3}$ | $4.5\times10^{-3}$ | $1.5\times10^{-3}$ |
| 20 | $5.5\times10^{-2}$ | $8.0\times10^{-3}$ | $4.7\times10^{-2}$ | $8.0\times10^{-3}$ | $8.0\times10^{-3}$ | 0 | $2.0\times10^{-5}$ | $2.0\times10^{-5}$ | 0 |
| 30 | $6.1\times10^{-3}$ | $8.2\times10^{-4}$ | $5.3\times10^{-3}$ | $8.2\times10^{-4}$ | $8.2\times10^{-4}$ | 0 | $1.0\times10^{-5}$ | $1.0\times10^{-5}$ | 0 |

## V. CONCLUSIONS

In this paper, we have investigated only the BER performance gap between the sub-optimal and optimal linear precoding schemes for a multiuser MIMO systems. Most prior works has been focused on the BER performance using the independent LP technique so that the performance gap is still remaining. Thus, we have well-designed the ULP scheme to mitigate completely this problem. The simulation results verify that the designed ULP scheme outperforms of the LP scheme and is able to achieve much better than BER performance. In future, it will be helpful to construct a unified transceiver for the next generation communication systems.